\documentclass[preprintnumbers,hyperref]{revtex4}
\usepackage{amssymb}
\usepackage{float}
\usepackage{graphics}
\usepackage{amsfonts}
\usepackage{amsmath}

\begin{document}
\title{Quantum Mechanical Perspective and Generalization of the Fractional
Fourier Transformation\footnotetext[1]{Work supported by the National Basic Research Program of China (973 Program,
2012CB922001), the National Natural Science Foundation of China under grant
11105133 and 11751113}\\
}

\author{Jun-Hua Chen$^{1,2,3,4}$ \footnotetext[2]{Corresponding Author, Email: cjh@ustc.edu.cn} and Hong-Yi Fan$^{3}$}
\address{
$^{1}$ {\small {Hefei Center for Physical Science and Technology, Hefei,
Anhui, 230026, China}}\\
$^{2}$ {\small {Synergetic Innovation Center of Quantum Information and
Quantum Physics}}\\
{\small { USTC, Hefei, Anhui, 230026, China}}\\
$^{3}$ {\small {Department of Material Science and Engineering, USTC, Hefei,
Anhui, 230026, China}}\\
$^{4}$ {\small {CAS Key Laboratory of Materials for Energy
Conversion, Hefei, Anhui, 230026, China}}}

\begin{abstract}
Fourier and fractional-Fourier transformations are widely used in
theoretical physics. In this paper we make quantum perspectives and
generalization for the fractional Fourier transformation (FrFT). By virtue
of quantum mechanical representation transformation and the method of
integration within normal ordered product (IWOP) of operators, we find the
key point for composing FrFT, and reveal the structure of FrFT. Following
this procedure, a full family of generalized fractional transformations are
discovered with the usual FrFT as one special case. The eigen-functions of
arbitrary GFrT are derived explicitly.\newline

\textbf{Keywords: }{fractional Fourier transformation; additivity; Abelian
Lie group; Parseval; eigen-function}
\end{abstract}

\maketitle

\section{Introduction}

The fractional Fourier transformation (FrFT) is a very useful tool in
Fourier optics and information optics, especially in optical communication,
image manipulations and signal analysis \cite{1,2,3,4,5,6,7}. The concept of
the FrFT was originally described by Condon \cite{3} and was later
introduced for signal processing in 1980 by Namias \cite{4} as a Fourier
transform of fractional order. Sumiyoshi et al also made an interesting
generalization on FrFT in 1994 \cite{8}. Another generalization was made by Shutian Liu et al in 1997\cite{9}. FrFT did not have significant
impact on optics until FrFT was defined physically based on propagation in
quadratic graded-index media (GRIN media). Mendlovic and Ozaktas \cite{5}
defined the $\alpha $-th FrFT as follows: let the original function be the
input at one side of quadratic GRIN medium at $z=0$, then the light
distribution observed on the plane $z=z_{0}$ equals to the ($z_{0}/L$)-th
fractional Fourier transform of the input function, where $L\equiv (\pi
/2)(n_{1}/n_{2})^{1/2}$ is the characteristic distance, $n_{1},n_{2}$ are
medium's physical parameters involved in the refractive index $%
n(r)=n_{1}-n_{2}r^{2}/2$, $r$ is the radial distance from the optical $z$
axis).

For real parameter $\alpha $, the 1-dimensional $\alpha $-angle FrFT of a
function $f$ is denoted by $F_{\alpha }\left[ f\right] $ and defined by%
\begin{equation}
F_{\alpha }\left[ f\right] \left( p\right) =\sqrt{\frac{1-i\cot \alpha }{%
2\pi }}\int_{-\infty }^{\infty }\exp \left[ \frac{i}{2}\left( \frac{%
p^{2}+x^{2}}{\tan \alpha }-\frac{2px}{\sin \alpha }\right) \right] f\left(
x\right) dx,  \label{1}
\end{equation}%
where the square root $\sqrt{1-i\cot \alpha }$ is defined such that $\arg
\sqrt{1-i\cot \alpha }$ $\in \left[ \pi /2,-\pi /2\right) $. The
conventional Fourier transform is simply $F_{\pi /2}$. The composition $%
F_{\alpha }\circ F_{\beta }$ of two FrFT's with parameters $\alpha $ and $%
\beta $ is defined by
\begin{equation}
\left( F_{\alpha }\circ F_{\beta }\right) \left[ f\right] \equiv F_{\alpha }%
\left[ F_{\beta }\left[ f\right] \right] .  \label{2}
\end{equation}%
$F$ is additive under definition Eq. (\ref{2}), i.e.,%
\begin{equation}
F_{\alpha }\circ F_{\beta }=F_{\alpha +\beta }.  \label{3}
\end{equation}%
In the context of quantum mechanics, function $f$ turns to quantum state $%
\left\vert f\right\rangle $, the value $f\left( x\right) $\ of $f$ at given
point $x$ turns to the matrix element $\left\langle x\right. \left\vert
f\right\rangle $ under coordinate representation $\left\vert x\right\rangle $%
. The usual Fourier transform is simply changing of basis to momentum
representation $\left\vert p\right\rangle $,%
\begin{eqnarray}
\hat{f}\left( p\right) &=&\left\langle p\right. \left\vert f\right\rangle
=\int_{-\infty }^{\infty }\left\langle p\right. \left\vert x\right\rangle
\left\langle x\right. \left\vert f\right\rangle dx  \label{4} \\
&=&\int_{-\infty }^{\infty }\frac{e^{-ipx}}{\sqrt{2\pi }}f\left( x\right) dx.
\notag
\end{eqnarray}%
And the $\alpha $-angle fractional Fourier transform is simply

\begin{equation}
F_{\alpha }\left[ f\right] \left( p\right) =\left\langle p\right\vert
e^{i\left( \frac{\pi }{2}-\alpha \right) a^{\dag }a}\left\vert
f\right\rangle =\int_{-\infty }^{\infty }K_{\alpha }\left( p,x\right)
f\left( x\right) dx,  \label{5}
\end{equation}%
where $a$ and $a^{\dag }$ are the annihilation and creation operator
respectively. The kernel of transformation $K_{\alpha }\left( p,x\right) $ is%
\begin{eqnarray}
K_{\alpha }\left( p,x\right) &=&\left\langle p\right\vert K_{\alpha
}\left\vert x\right\rangle =\left\langle p\right\vert e^{i\left( \frac{\pi }{%
2}-\alpha \right) a^{\dag }a}\left\vert x\right\rangle  \label{6} \\
&=&\sqrt{\frac{1-i\cot \alpha }{2\pi }}\exp \left[ \frac{i}{2}\left( \frac{%
p^{2}+x^{2}}{\tan \alpha }-\frac{2px}{\sin \alpha }\right) \right] .  \notag
\end{eqnarray}%
I.e., the $\alpha $-angle fractional Fourier transform is the composite
transformation of both the basis changing and unitary transformation
generated by the operator $K_{\alpha }=e^{i\left( \frac{\pi }{2}-\alpha
\right) a^{\dag }a}$. The key feature here is\ that the transformation is
compositable and additive, i.e., one can perform fractional Fourier
transform repeatedly on given function $f$,

\begin{equation}
\left( F_{\alpha }\circ F_{\beta }\right) \left[ f\right] \equiv F_{\alpha }%
\left[ F_{\beta }\left[ f\right] \right] ,  \label{7}
\end{equation}%
and $F$ is additive under definition Eq. (\ref{7}), $F_{\alpha }\circ
F_{\beta }=F_{\alpha +\beta }$.

Enlightened by the above analysis, we hope to find the criteria for
constructing new generalized fractional transformation (GFrT). In other
words, we want to generalize FrFT to all possible compositable and additive
transformations which automatically exhibit fractional transform's
properties. We shall do this by virtue of quantum mechanical representation
transformation \cite{10} and the method of integration within normal ordered
product (IWOP) of operators \cite{11}.

\section{Analysis of the Key Point of GFrT}

Let $\left\{ \left\vert \boldsymbol{A}\right\rangle \text{'s}\right\} $ and $%
\left\{ \left\vert \boldsymbol{B}\right\rangle \text{'s}\right\} $ denote
two sets of basis. In order to perform their mutual transformation
repeatedly, $\left\vert \boldsymbol{A}\right\rangle $ and $\left\vert
\boldsymbol{B}\right\rangle $ must have matching parameterizations, i.e.,
parameters $\boldsymbol{A}\boldsymbol{=}\left( A_{1},\cdots ,A_{n}\right) $
and $\boldsymbol{B}\boldsymbol{=}\left( B_{1},\cdots ,B_{n}\right) $ are in
the same Borel set $\mathbb{D}$. $\mathbb{D}$\ is assigned with measure $\mu
$ so that a proper Lebesgue integration can be defined on $\mathbb{D}$. And

As usual, we demand the completeness of $\left\vert \boldsymbol{A}%
\right\rangle $ and $\left\vert \boldsymbol{B}\right\rangle $
\begin{equation}
\int \left\vert \boldsymbol{A}\right\rangle \left\langle \boldsymbol{A}%
\right\vert d\mu \left( \boldsymbol{A}\right) \boldsymbol{=1,}\text{ }\int
\left\vert \boldsymbol{B}\right\rangle \left\langle \boldsymbol{B}%
\right\vert d\mu \left( \boldsymbol{B}\right) \boldsymbol{=1}  \label{9}
\end{equation}%
General transformation $F_{K}$ on function $f$ of $\boldsymbol{A}$ is
defined by
\begin{eqnarray}
F_{\hat{K}}\left[ f\right] \left( \boldsymbol{B}\right) &=&\left\langle
\boldsymbol{B}\right\vert \hat{K}\left\vert f\right\rangle =\int
\left\langle \boldsymbol{B}\right\vert \hat{K}\left\vert \boldsymbol{A}%
\right\rangle \left\langle \boldsymbol{A}\right. \left\vert f\right\rangle
d\mu \left( \boldsymbol{A}\right)  \label{11} \\
&=&\int K\left( \boldsymbol{B},\boldsymbol{A}\right) \left\langle
\boldsymbol{A}\right. \left\vert f\right\rangle d\mu \left( \boldsymbol{A}%
\right) ,  \notag
\end{eqnarray}%
where $K\left( \boldsymbol{B},\boldsymbol{A}\right) $ is a composite
\begin{equation}
K\left( \boldsymbol{B},\boldsymbol{A}\right) \equiv \left\langle \boldsymbol{%
B}\right\vert \hat{K}\left\vert \boldsymbol{A}\right\rangle .  \label{12}
\end{equation}%
Since $F_{K}$ is a transformation on functions, and functions $f$ and $F_{%
\hat{K}}\left[ f\right] $ have matching variables, it is natural to define
the composite transformation $F_{\hat{K}_{1}}\circ F_{\hat{K}_{2}}$ of $F_{%
\hat{K}_{1}}$ and $F_{\hat{K}_{2}}$ on function $f$ by
\begin{equation}
\left( F_{\hat{K}_{1}}\circ F_{\hat{K}_{2}}\right) \left[ f\right] =F_{\hat{K%
}_{1}}\left[ F_{\hat{K}_{2}}\left[ f\right] \right] ,  \label{13}
\end{equation}%
i.e.
\begin{equation}
\left( F_{\hat{K}_{1}}\circ F_{\hat{K}_{2}}\right) \left[ f\right] \left(
\boldsymbol{B}\right) =\int \int K_{1}\left( \boldsymbol{B},\boldsymbol{A}%
\right) K_{2}\left( \boldsymbol{A},\boldsymbol{A}^{\prime }\right) f\left(
\boldsymbol{A}^{\prime }\right) d\mu \left( \boldsymbol{A}^{\prime }\right)
d\mu \left( \boldsymbol{A}\right) ,  \label{14}
\end{equation}%
where, according to Eq. (\ref{12}),
\begin{equation}
K_{1}\left( \boldsymbol{B},\boldsymbol{A}\right) =\left\langle \boldsymbol{B}%
\right\vert \hat{K}_{1}\left\vert \boldsymbol{A}\right\rangle ,  \label{15}
\end{equation}%
and $K_{2}\left( \boldsymbol{A},\boldsymbol{A}^{\prime }\right) $ should be
composite too,%
\begin{equation}
K_{2}\left( \boldsymbol{A},\boldsymbol{A}^{\prime }\right) =\left\langle
\boldsymbol{B}^{\prime }\right\vert _{\boldsymbol{B}^{\prime }\boldsymbol{=A}%
}\hat{K}_{2}\left\vert \boldsymbol{A}^{\prime }\right\rangle .  \label{16}
\end{equation}%
Here $\left\vert \boldsymbol{B}^{\prime }\right\rangle _{\boldsymbol{B}%
^{\prime }\boldsymbol{=A}}$ means a $\left\vert \boldsymbol{B}^{\prime
}\right\rangle $ state with parameter $\boldsymbol{B}^{\prime }\boldsymbol{=A%
}$, for example, $\left\vert p^{\prime }\right\rangle _{p^{\prime }=x}$ is a
momentum eigenstate with eigenvalue $p^{\prime }=x$. Eq. (\ref{14}) then
reads%
\begin{eqnarray}
&&\left( F_{\hat{K}_{1}}\circ F_{\hat{K}_{2}}\right) \left[ f\right] \left(
\boldsymbol{B}\right)  \label{17} \\
&=&\int \int \left\langle \boldsymbol{B}\right\vert \hat{K}_{1}\left\vert
\boldsymbol{A}\right\rangle \left\langle \boldsymbol{B}^{\prime }\right\vert
_{\boldsymbol{B}^{\prime }\boldsymbol{=A}}\hat{K}_{2}\left\vert \boldsymbol{A%
}^{\prime }\right\rangle \left\langle \boldsymbol{A}^{\prime }\right.
\left\vert f\right\rangle d\mu \left( \boldsymbol{A}^{\prime }\right) d\mu
\left( \boldsymbol{A}\right)  \notag \\
&=&\left\langle \boldsymbol{B}\right\vert \hat{K}_{1}\mathfrak{M}\hat{K}%
_{2}\left\vert \boldsymbol{f}\right\rangle =F_{\hat{K}_{1}\mathfrak{M}\hat{K}%
_{2}}\left[ f\right] \left( \boldsymbol{B}\right) ,  \notag
\end{eqnarray}%
where an operator $\mathfrak{M}$ emerged%
\begin{equation}
\mathfrak{M}\equiv \int \left\vert \boldsymbol{A}\right\rangle \left\langle
\boldsymbol{B}^{\prime }\right\vert _{\boldsymbol{B}^{\prime }\boldsymbol{=A}%
}d\mu \left( \boldsymbol{A}\right) .  \label{18}
\end{equation}%
This operator is an essence for the successive composite transformations.
Note that the rule of composition of transformations is%
\begin{equation}
F_{\hat{K}_{1}}\circ F_{\hat{K}_{2}}=F_{\hat{K}_{1}\mathfrak{M}\hat{K}%
_{2}}\neq F_{\hat{K}_{1}\hat{K}_{2}},  \label{19}
\end{equation}%
i.e., the mapping $\hat{K}\rightarrow F_{\hat{K}}$ is not homomorphic. The
key point of defining and finding GFrT is to determine all allowed operators
$\hat{K}$ so that the transformations are compositable and additive.

\section{Determination of Allowed $\hat{K}$}

We determine the allowed operator $\hat{K}$ by two criteria, the first is
the additivity of the transformations, the second is that the
transformations must also satisfy the Parseval theorem as the conventional
Fourier transformation obeys.

\subsection{Additivity}

Since the fundamental property of FrFT is the additivity, $F_{\alpha }\circ
F_{\beta }=F_{\alpha +\beta }$, we demand that the additivity still holds
for GFrT $F_{\hat{K}}$, therefore there must be a way of parameterization $%
\hat{K}_{\alpha }$ for $\hat{K}$ such that
\begin{equation}
\hat{K}_{\alpha }\mathfrak{M}\hat{K}_{\beta }=\hat{K}_{\alpha +\beta }.
\label{20}
\end{equation}%
This equation is necessary and sufficient for $F_{\hat{K}}$ to be additive,%
\begin{equation}
F_{\hat{K}_{\alpha }}\circ F_{\hat{K}_{\beta }}=F_{\hat{K}_{\alpha }%
\mathfrak{M}\hat{K}_{\beta }}=F_{\hat{K}_{\alpha +\beta }}.  \label{21}
\end{equation}%
In this case $F_{\hat{K}_{\alpha }}$ can be considered as the natural
generalization of the FrFT.

By defining
\begin{equation}
\tilde{K}_{\alpha }=\hat{K}_{\alpha }\mathfrak{M},  \label{22}
\end{equation}%
equation $\hat{K}_{\alpha }\mathfrak{M}\hat{K}_{\beta }=\hat{K}_{\alpha
+\beta }$ becomes%
\begin{equation}
\tilde{K}_{\alpha }\tilde{K}_{\beta }=\tilde{K}_{\alpha +\beta },  \label{23}
\end{equation}%
i.e. the allowed $\tilde{K}_{\alpha }$'s form an Abelian Lie group $\mathcal{%
\tilde{K}}$ (or the subgroup of an Abelian Lie group, in the case that some
components of $\alpha $\ take discrete values), $\hat{K}_{\alpha }$'s form
the right coset $\mathcal{\tilde{K}}\mathfrak{M}^{-1}$ of $\mathcal{\tilde{K}%
}$. Conversely, if we have an Abelian Lie group $\mathcal{\tilde{K}}$ and
two sets of basis $\left\vert \boldsymbol{A}\right\rangle $ and $\left\vert
\boldsymbol{B}\right\rangle $ with matching parameterization, then we can
define a GFrT by

\begin{equation}
F_{\hat{K}_{\alpha }}\left[ f\right] \left( \boldsymbol{B}\right)
=\left\langle \boldsymbol{B}\right\vert \hat{K}_{\alpha }\left\vert
f\right\rangle =\int K_{\alpha }\left( \boldsymbol{B},\boldsymbol{A}\right)
\left\langle \boldsymbol{A}\right. \left\vert f\right\rangle d\mu \left(
\boldsymbol{A}\right)  \label{24}
\end{equation}%
where $\hat{K}_{\alpha }=\tilde{K}_{\alpha }\mathfrak{M}^{-1}$.

\subsection{Parseval's Theorem}

Further, we demand some sort of Parseval's theorem for the new
transformation, i.e.

\begin{eqnarray}
\int \left\vert F_{\hat{K}}\left[ f\right] \left( \boldsymbol{B}\right)
\right\vert ^{2}d\mu \left( \boldsymbol{B}\right) &\equiv &\int \left\vert
f\left( \boldsymbol{A}\right) \right\vert ^{2}d\mu \left( \boldsymbol{A}%
\right)  \label{25} \\
\int \left\langle f\right\vert \hat{K}^{\dag }\left\vert \boldsymbol{B}%
\right\rangle \left\langle \boldsymbol{B}\right\vert \hat{K}\left\vert
f\right\rangle d\mu \left( \boldsymbol{B}\right) &\equiv &\int \left\langle
f\right. \left\vert \boldsymbol{A}\right\rangle \left\langle \boldsymbol{A}%
\right. \left\vert f\right\rangle d\mu \left( \boldsymbol{A}\right)  \notag
\\
\left\langle f\right\vert \hat{K}^{\dag }\hat{K}\left\vert f\right\rangle
&\equiv &\left\langle f\right. \left\vert f\right\rangle  \notag
\end{eqnarray}%
therefore $\hat{K}$ must be unitary, $\hat{K}^{\dag }\hat{K}=1$.

Parseval's theorem demands that all the allowed $\hat{K}^{\prime }s$ must be
unitary, therefore $\hat{K}_{1}\mathfrak{M}\hat{K}_{2}$ must be unitary if
we wish to define $F_{\hat{K}_{1}}\circ F_{\hat{K}_{2}}$ properly. Operator $%
\mathfrak{M=}\hat{K}_{1}^{-1}\left( \hat{K}_{1}\mathfrak{M}\hat{K}%
_{2}\right) \hat{K}_{2}^{-1}$ must be unitary too.

The unitarity of $\mathfrak{M}$ is guaranteed if either $\left\vert
\boldsymbol{A}\right\rangle $'s or $\left\vert \boldsymbol{B}\right\rangle $%
's are orthonormal. In fact, when $\left\langle \boldsymbol{A}^{\prime
}\right. \left\vert \boldsymbol{A}\right\rangle =\delta ^{\left( n\right)
}\left( \boldsymbol{A}^{\prime }-\boldsymbol{A}\right) $, i.e., $\left\vert
\boldsymbol{A}\right\rangle $'s are orthonormal, then%
\begin{eqnarray}
\mathfrak{M}^{\dag }\mathfrak{M} &\mathfrak{=}&\int \left\vert \boldsymbol{B}%
^{\prime }\right\rangle _{\boldsymbol{B}^{\prime }\boldsymbol{=A}^{\prime
}}\left\langle \boldsymbol{A}^{\prime }\right\vert d^{n}\boldsymbol{A}%
^{\prime }\int \left\vert \boldsymbol{A}\right\rangle \left\langle
\boldsymbol{B}\right\vert _{\boldsymbol{B=A}}d\mu \left( \boldsymbol{A}%
\right)  \label{27} \\
&=&\int \left\vert \boldsymbol{B}^{\prime }\right\rangle _{\boldsymbol{B}%
^{\prime }\boldsymbol{=A}^{\prime }}\left\langle \boldsymbol{B}\right\vert _{%
\boldsymbol{B=A}}\delta ^{\left( n\right) }\left( \boldsymbol{A}^{\prime }-%
\boldsymbol{A}\right) d\mu \left( \boldsymbol{A}^{\prime }\right) d\mu
\left( \boldsymbol{A}\right)  \notag \\
&=&\int \left\vert \boldsymbol{B}^{\prime }\right\rangle _{\boldsymbol{B}%
^{\prime }\boldsymbol{=A}}\left\langle \boldsymbol{B}\right\vert _{%
\boldsymbol{B=A}}d\mu \left( \boldsymbol{A}\right) =1.  \notag
\end{eqnarray}%
If $\left\langle \boldsymbol{B}^{\prime }\right. \left\vert \boldsymbol{B}%
\right\rangle =\delta ^{\left( n\right) }\left( \boldsymbol{B}^{\prime }-%
\boldsymbol{B}\right) $, then we also have%
\begin{eqnarray}
\mathfrak{MM}^{\dag } &\equiv &\int \left\vert \boldsymbol{A}\right\rangle
\left\langle \boldsymbol{B}^{\prime }\right\vert _{\boldsymbol{B}^{\prime }%
\boldsymbol{=A}}d^{n}\boldsymbol{A}\int \left\vert \boldsymbol{B}^{\prime
}\right\rangle _{\boldsymbol{B}^{\prime }\boldsymbol{=A}^{\prime
}}\left\langle \boldsymbol{A}^{\prime }\right\vert d\mu \left( \boldsymbol{A}%
^{\prime }\right)  \label{28} \\
&=&\int \left\vert \boldsymbol{A}\right\rangle \left\langle \boldsymbol{A}%
^{\prime }\right\vert \delta ^{\left( n\right) }\left( \boldsymbol{B}%
^{\prime }-\boldsymbol{B}\right) |_{\boldsymbol{B}^{\prime }\boldsymbol{=A,B}%
^{\prime }\boldsymbol{=A}^{\prime }}d\mu \left( \boldsymbol{A}^{\prime
}\right) d\mu \left( \boldsymbol{A}\right)  \notag \\
&=&\int \left\vert \boldsymbol{A}\right\rangle \left\langle \boldsymbol{A}%
\right\vert d\mu \left( \boldsymbol{A}\right) =1.  \notag
\end{eqnarray}%
$F_{\hat{K}_{\alpha }}$'s defined by Eq. (\ref{24}) for such $\mathcal{%
\tilde{K}}$ and $\left\vert \boldsymbol{A}\right\rangle $, $\left\vert
\boldsymbol{B}\right\rangle $ are additive and satisfy Parseval's Theorem.

\section{The Construction and the Eigen-problem of GFrT}

We need a unitary Abel Lie group $\mathcal{\tilde{K}}$ to construct the
generalized transformation satisfying Parseval's Theorem. The structure of
such group $\mathcal{\tilde{K}}$ is simple, each element takes the form $%
\tilde{K}_{\alpha }=\exp \left[ i\sum\limits_{j}\alpha _{j}\hat{O}_{j}\right]
$, where $\alpha _{j}$ are real parameters and $\hat{O}_{j}$ are Hermitian
operators that commute with each other, $\left[ \hat{O}_{j},\hat{O}_{k}%
\right] =0$. Conversely, given a set of commutating Hermitian operators $%
\hat{O}_{j}$(there is no other constraints on $\hat{O}_{j}$ other than
Hermiticity) and two sets of basis $\left\vert \boldsymbol{A}\right\rangle $
and $\left\vert \boldsymbol{B}\right\rangle $ with matching
parameterization, we can construct the corresponding generalized fractional
transform $F_{\hat{K}_{\alpha }}$ with
\begin{equation}
\hat{K}_{\alpha }=\tilde{K}_{\alpha }\mathfrak{M}^{\dag }=\exp \left[
i\sum\limits_{j}\alpha _{j}\hat{O}_{j}\right] \mathfrak{M}^{\dag }.
\label{29}
\end{equation}

Now it is obvious that any transformation $F$ on functions that takes the
form $F\left[ f\right] \left( \boldsymbol{B}\right) =\int K\left(
\boldsymbol{B},\boldsymbol{A}\right) f\left( \boldsymbol{A}\right) d\mu
\left( \boldsymbol{A}\right) $ and leaves the domains of the functions
unchanged can be extended to GFrT through the following standard procedure.
First we find two basis $\left\vert \boldsymbol{A}\right\rangle $ and $%
\left\vert \boldsymbol{B}\right\rangle $ with proper parameterizations ($%
\left\vert \boldsymbol{B}\right\rangle $ can be chosen to be $\left\vert
\boldsymbol{A}\right\rangle $ if one does not care about the
\textquotedblleft physical interpretation" of the GFrT), then define
operators $\hat{K}_{1}=\int \int K\left( \boldsymbol{B},\boldsymbol{A}%
\right) \left\vert \boldsymbol{B}\right\rangle \left\langle \boldsymbol{A}%
\right\vert d\mu \left( \boldsymbol{A}\right) d\mu \left( \boldsymbol{B}%
\right) $ and $\mathfrak{M}\equiv \int \left\vert \boldsymbol{A}%
\right\rangle \left\langle \boldsymbol{B}^{\prime }\right\vert _{\boldsymbol{%
B}^{\prime }\boldsymbol{=A}}d\mu \left( \boldsymbol{A}\right) $. The GFrT $%
F_{\alpha }$ generated by $\hat{K}_{\alpha }=\left( K_{1}\mathfrak{M}\right)
^{\alpha }\mathfrak{M}^{-1}$ is the natural extension of $F$ and $F_{1}=F$.

The eigen-problem is one of the most important objects for each linear transformation. For example. the eigen-problem of the fractional Fourier transformation was discussed in \cite{12} by Alieva et al. Since we have found a better perspective of the fractional transformations, here the eigen-problem can be solved more neatly and generally. As usual, the eigen-functions $f$ of classical GFrT must satisfy

\begin{equation}
\hat{f}\left( \boldsymbol{B}\right) =\int K\left( \boldsymbol{B},\boldsymbol{%
A}\right) f\left( \boldsymbol{A}\right) d\mu \left( \boldsymbol{A}\right)
=\lambda f\left( \boldsymbol{B}\right) .  \label{30}
\end{equation}%
The quantum version of Eq. (\ref{30}) is%
\begin{equation}
\left\langle \boldsymbol{B}\right\vert \left. \hat{f}\right\rangle
=\left\langle \boldsymbol{B}\right\vert \hat{K}\left\vert f\right\rangle
=\lambda \left\langle \boldsymbol{A}\right\vert \left. f\right\rangle _{%
\boldsymbol{A}=\boldsymbol{B}},  \label{31}
\end{equation}%
which is equivalent to%
\begin{eqnarray}
\int \left\vert \boldsymbol{B}\right\rangle \left\langle \boldsymbol{B}%
\right\vert \hat{K}\left\vert f\right\rangle d\mu \left( \boldsymbol{B}%
\right) &=&\lambda \int \left\vert \boldsymbol{B}\right\rangle \left\langle
\boldsymbol{A}\right\vert \left. f\right\rangle _{\boldsymbol{A}=\boldsymbol{%
B}}d\mu \left( \boldsymbol{B}\right)  \label{32} \\
\hat{K}\left\vert f\right\rangle &=&\lambda \mathfrak{M}^{\dag }\left\vert
f\right\rangle  \notag \\
\exp \left[ \sum_{j=1}^{s}i\alpha _{j}\hat{O}_{j}\right] \mathfrak{M}^{\dag
}\left\vert f\right\rangle &=&\lambda \mathfrak{M}^{\dag }\left\vert
f\right\rangle .  \label{33}
\end{eqnarray}

Since $\hat{O}_{j}$'s commute with each other, eigen-equation Eq. (\ref{33})
can be decomposed as equations%
\begin{equation}
\hat{O}_{j}\mathfrak{M}^{\dag }\left\vert f\right\rangle =\theta _{j}%
\mathfrak{M}^{\dag }\left\vert f\right\rangle ,  \label{34}
\end{equation}%
i.e., $\mathfrak{M}^{\dag }\left\vert f\right\rangle $ is the common
eigenstate of the commuting Hermitian operators $\hat{O}_{j}$'s. Let $%
\left\vert \varphi _{m}\right\rangle $ be the common eigenstate of $\hat{O}%
_{j}$'s%
\begin{equation}
\hat{O}_{j}\left\vert \varphi _{m}\right\rangle =\theta _{j,m}\left\vert
\varphi _{m}\right\rangle  \label{35}
\end{equation}%
($\left\vert \varphi _{m}\right\rangle $'s form a complete set, $%
\sum_{m}\left\vert \varphi _{m}\right\rangle \left\langle \varphi
_{m}\right\vert =1$, and $\left\vert \varphi _{m}\right\rangle $ can be
chosen to be orthogonal, $\left\langle \varphi _{m}\right\vert \left.
\varphi _{m^{\prime }}\right\rangle =\delta _{m,m^{\prime }}$), then
\begin{equation}
\left\vert f_{m}\right\rangle =\mathfrak{M}\left\vert \varphi
_{m}\right\rangle .  \label{36}
\end{equation}%
The eigen-functions of the classical GFrT is%
\begin{equation}
f_{m}\left( \boldsymbol{A}\right) =\left\langle \boldsymbol{A}\right\vert
\left. f_{m}\right\rangle =\left\langle \boldsymbol{A}\right\vert \mathfrak{M%
}\left\vert \varphi _{m}\right\rangle  \label{37}
\end{equation}%
with eigenvalue $\exp \left[ \sum_{j=1}^{s}i\alpha _{j}\theta _{j,m}\right] $%
,
\begin{equation}
F_{K_{\alpha }}\left[ f_{m}\right] \left( \boldsymbol{B}\right) =\exp \left[
\sum_{j=1}^{s}i\alpha _{j}\theta _{j,m}\right] f_{m}\left( \boldsymbol{B}%
\right)  \label{38}
\end{equation}%
The eigen-functions $f_{m}\left( \boldsymbol{A}\right) $'s are orthogonal%
\begin{eqnarray}
\int f_{m}^{\ast }\left( \boldsymbol{A}\right) f_{m^{\prime }}\left(
\boldsymbol{A}\right) d\mu \left( \boldsymbol{A}\right) &=&\int \left\langle
\varphi _{m}\right\vert \mathfrak{M}^{\dag }\left\vert \boldsymbol{A}%
\right\rangle \left\langle \boldsymbol{A}\right\vert \mathfrak{M}\left\vert
\varphi _{m^{\prime }}\right\rangle d\mu \left( \boldsymbol{A}\right)
\label{39} \\
&=&\left\langle \varphi _{m}\right\vert \mathfrak{M}^{\dag }\mathfrak{M}%
\left\vert \varphi _{m^{\prime }}\right\rangle =\delta _{m,m^{\prime }}.
\notag
\end{eqnarray}%
And $f_{m}\left( \boldsymbol{A}\right) $'s are complete too, any
\textquotedblleft good" functions $g\left( \boldsymbol{A}\right) $ can be
written as the linear combination of $f_{m}\left( \boldsymbol{A}\right) $'s,
\begin{eqnarray}
g\left( \boldsymbol{A}\right) &=&\left\langle \boldsymbol{A}\right\vert
\left. g\right\rangle =\sum_{m}\left\langle \boldsymbol{A}\right\vert
\mathfrak{M}\left\vert \varphi _{m}\right\rangle \left\langle \varphi
_{m}\right\vert \mathfrak{M}^{\dag }\left\vert g\right\rangle  \label{40} \\
&=&\sum_{m}\left\langle f_{m}\right\vert \left. g\right\rangle \left\langle
\boldsymbol{A}\right\vert \left. f_{m}\right\rangle
=\sum_{m}C_{m}f_{m}\left( \boldsymbol{A}\right) .  \notag
\end{eqnarray}

.

\section{Some Examples}

\subsection{Example 1}

Let $\left\vert \boldsymbol{A}\right\rangle =\left\vert x\right\rangle $,
the coordinate representation, and $\left\vert \boldsymbol{B}\right\rangle
=\left\vert p\right\rangle $, the momentum representation, in Fock space
they are expressed as
\begin{eqnarray}
\left\vert \boldsymbol{A}\right\rangle &=&\left\vert x\right\rangle =\frac{1%
}{\pi ^{1/4}}\exp \left[ -\frac{1}{2}x^{2}+\sqrt{2}xa^{\dag }-\frac{1}{2}%
a^{\dag 2}\right] \left\vert 0\right\rangle ,  \label{41} \\
\left\vert \boldsymbol{B}\right\rangle &=&\left\vert p\right\rangle =\frac{1%
}{\pi ^{1/4}}\exp \left[ -\frac{1}{2}p^{2}+i\sqrt{2}pa^{\dag }+\frac{1}{2}%
a^{\dag 2}\right] \left\vert 0\right\rangle  \notag
\end{eqnarray}%
where $\left[ a,a^{\dag }\right] =1,$ and $\left\vert 0\right\rangle $ is
the vacuum state annihilated by $a$, $a\left\vert 0\right\rangle =0$. Using
the method of integration within normal ordered product of operators \cite{11}
and the fact that $\left\vert 0\right\rangle \left\langle 0\right\vert
=:\exp [-a^{\dag }a]:$, we can perform the following integration
(constructed according to Eq. (\ref{18}))
\begin{eqnarray}
\mathfrak{M} &=&\int_{-\infty }^{\infty }dx\left\vert x\right\rangle
\left\langle p\right\vert _{p=x}  \label{42} \\
&=&\int_{-\infty }^{\infty }\frac{dx}{\sqrt{\pi }}:\exp \left[ -x^{2}+\sqrt{2%
}xa^{\dag }-\frac{1}{2}a^{\dag 2}-a^{\dag }a-i\sqrt{2}xa+\frac{1}{2}a^{2}%
\right] :  \notag \\
&=&:\exp \left[ -\left( 1+i\right) a^{\dag }a\right] :=\exp \left[ -\frac{%
i\pi }{2}a^{\dag }a\right] ,  \notag
\end{eqnarray}%
obviously $\hat{K}_{\alpha }=e^{i\left( \frac{\pi }{2}-\alpha \right)
a^{\dag }a}$ in Eq. (\ref{6}) obeys $\hat{K}_{\alpha }\mathfrak{M}\hat{K}%
_{\beta }=\hat{K}_{\alpha +\beta }$, no wonder Eqs. (\ref{5})-(\ref{6}) can
embody the characters of fractional Fourier transform. And we now understand
better why the kernel $K_{\alpha }\left( p,x\right) $ should be defined in
the seemingly unnatural way $\left\langle p\right\vert e^{i\left( \frac{\pi
}{2}-\alpha \right) a^{\dag }a}\left\vert x\right\rangle $ instead of
naturally $\left\langle p\right\vert e^{-i\alpha a^{\dag }a}\left\vert
x\right\rangle $, this is because $\hat{K}_{\alpha }=\tilde{K}_{\alpha }%
\mathfrak{M}^{\dag }=e^{-i\alpha a^{\dag }a}e^{i\frac{\pi }{2}a^{\dag
}a}=e^{i\left( \frac{\pi }{2}-\alpha \right) a^{\dag }a}$. And we see that
the eigen-functions of FrFT are

\begin{eqnarray}
f_{m}\left( x\right) &=&\left\langle x\right\vert \mathfrak{M}\left\vert
m\right\rangle =\left\langle x\right\vert \exp \left[ -\frac{i\pi }{2}%
a^{\dag }a\right] \left\vert m\right\rangle  \label{43} \\
&=&\frac{1}{i^{m}}\left\langle x\right\vert \left. m\right\rangle =\frac{1}{%
i^{m}\sqrt{2^{m}m!}}H_{m}\left( x\right) e^{-x^{2}/2}  \notag
\end{eqnarray}%
with eigenvalues $e^{-im\alpha }$ by Eqn. (\ref{37}).

\subsection{Example 2}

Let $\left\vert \boldsymbol{A}\right\rangle =\left\vert x\right\rangle $, $%
\left\vert \boldsymbol{B}\right\rangle =\left\vert p\right\rangle $, and

\begin{equation}
\tilde{K}_{\alpha }=\exp \left[ -\frac{i\alpha }{2}\left( a^{2}e^{i\theta
}+e^{-i\theta }a^{\dag 2}\right) \right] ,  \label{44}
\end{equation}%
$\mathcal{\tilde{K}}$ is Abelian with respect to the parameter $\alpha $.
The disentangling of $\tilde{K}_{\alpha }$ is%
\begin{equation}
\tilde{K}_{\alpha }=\frac{1}{\sqrt{\cosh \alpha }}:\exp \left[ -\frac{i}{2}%
a^{\dag 2}e^{-i\theta }\tanh \alpha +a^{\dag }a\left( \frac{1}{\cosh \alpha }%
-1\right) -\frac{i}{2}a^{2}e^{i\theta }\tanh \alpha \right] :.  \label{45}
\end{equation}%
Using the completeness relation of the coherent state%
\begin{equation}
\int \frac{d^{2}z}{\pi }\left\vert z\right\rangle \left\langle z\right\vert
=1,\text{ }\left\vert z\right\rangle =e^{-|z|^{2}/2+za^{\dag }}\left\vert
0\right\rangle  \label{46}
\end{equation}%
and knowing $\mathfrak{M}$ in Eq. (\ref{42}) having the property $\mathfrak{M%
}^{\dag }\left\vert x\right\rangle =\left\vert p^{\prime }\right\rangle
_{p^{\prime }=x}$, we have%
\begin{equation}
\left\langle p\right\vert \tilde{K}_{\alpha }\mathfrak{M}^{\dag }\left\vert
x\right\rangle =\int \frac{d^{2}z_{1}}{\pi }\frac{d^{2}z_{2}}{\pi }%
\left\langle p\right\vert \left. z_{1}\right\rangle \left\langle
z_{1}\right\vert \tilde{K}_{\alpha }\left\vert z_{2}\right\rangle
\left\langle z_{2}\right. \left\vert p^{\prime }\right\rangle _{p^{\prime
}=x}.  \label{47}
\end{equation}%
Then using the overlap%
\begin{equation}
\left\langle p\right\vert \left. z\right\rangle =\frac{1}{\pi ^{1/4}}\exp %
\left[ -\frac{p^{2}}{2}-\frac{\left\vert z\right\vert ^{2}}{2}+\frac{z^{2}}{2%
}-\sqrt{2}ipz\right] ,  \label{48}
\end{equation}%
and the integration formula%
\begin{eqnarray}
&&\int \frac{d^{2}z}{\pi }\exp \left[ \zeta \left\vert z\right\vert
^{2}+fz^{2}+gz^{\ast 2}+\xi z+\eta z^{\ast }\right]  \label{49} \\
&=&\frac{1}{\sqrt{\zeta ^{2}-4fg}}\exp \left[ \frac{-\zeta \xi \eta +f\eta
^{2}+g\xi ^{2}}{\zeta ^{2}-4fg}\right] ,  \notag
\end{eqnarray}%
we see%
\begin{equation}
\begin{array}{c}
\left\langle p\right\vert \tilde{K}_{\alpha }\mathfrak{M}^{\dag }\left\vert
x\right\rangle =\frac{1}{\sqrt{\pi \cosh \alpha }}\int \frac{d^{2}z_{1}}{\pi
}\int \frac{d^{2}z_{2}}{\pi } \\
\times \exp [-\left\vert z_{1}\right\vert ^{2}+\frac{z_{1}^{2}}{2}-\frac{i}{2%
}z_{1}^{\ast 2}e^{-i\theta }\tanh \alpha -\sqrt{2}ipz_{1}+\frac{1}{\cosh
\alpha }z_{1}^{\ast }z_{2} \\
-\frac{x^{2}}{2}-\frac{p^{2}}{2}-\frac{i}{2}z_{2}^{2}e^{i\theta }\tanh
\alpha -\left\vert z_{2}\right\vert ^{2}+\frac{z_{2}^{\ast 2}}{2}+\sqrt{2}%
ixz_{2}] \\
=\frac{1}{\sqrt{2\pi i\cos \theta \sinh \alpha }}\exp \left[ \frac{i}{2}%
\left( \frac{x^{2}+p^{2}}{\tanh \alpha \cos \theta }+\left(
x^{2}-p^{2}\right) \tan \theta -\frac{2xp}{\sinh \alpha \cos \theta }\right) %
\right] .%
\end{array}
\label{50}
\end{equation}%
One can easily check that this new transformations satisfy
\begin{eqnarray}
&&\left( F_{\alpha }\circ F_{\beta }\right) \left[ f\right] \left( p\right) =%
\frac{1}{\sqrt{2\pi i\cos \theta \sinh \alpha }\sqrt{2\pi i\cos \theta \sinh
\beta }}  \label{51} \\
&&\times \int dq\exp \left[ \frac{i}{2}\left( \frac{q^{2}+p^{2}}{\tanh
\alpha \cos \theta }+\left( q^{2}-p^{2}\right) \tan \theta -\frac{2qp}{\sinh
\alpha \cos \theta }\right) \right]  \notag \\
&&\times \int dx\exp \left[ \frac{i}{2}\left( \frac{x^{2}+q^{2}}{\tanh \beta
\cos \theta }+\left( x^{2}-q^{2}\right) \tan \theta -\frac{2xq}{\sinh \beta
\cos \theta }\right) \right] f\left( x\right)  \notag \\
&=&F_{\alpha +\beta }\left[ f\right] \left( p\right) ,  \notag
\end{eqnarray}%
so the transformations are additive.

If we choose $\theta =\frac{\pi }{2}$ in $\tilde{K}_{\alpha }$, $\tilde{K}%
_{\alpha }$ becomes the single-mode squeezing operator $\exp \left[ \frac{%
\alpha }{2}\left( a^{2}-a^{\dag 2}\right) \right] =\int_{-\infty }^{\infty
}e^{a/2}dp\left\vert pe^{a}\right\rangle \left\langle p\right\vert
=\int_{-\infty }^{\infty }e^{-a/2}dx\left\vert xe^{-a}\right\rangle
\left\langle x\right\vert $, the corresponding transformation becomes the
Hadamard transformation of continuum variables \cite{13}, i.e.,
\begin{equation}
K_{\alpha }=\exp \left[ \frac{\alpha }{2}\left( a^{2}-a^{\dag 2}\right) %
\right] \exp \left[ \frac{i\pi }{2}a^{\dag }a\right]  \label{52}
\end{equation}%
we have%
\begin{equation}
\lim_{\theta \rightarrow \frac{\pi }{2}}K_{\alpha }\left( p,x\right)
=e^{-a/2}\delta \left( x-pe^{-a}\right)  \label{53}
\end{equation}%
Hadamard transform is not only an important tool in classical signal
processing, but also is of great importance for quantum computation
applications .

If we choose $\theta =0$ in $\tilde{K}_{\alpha }$, $\tilde{K}_{\alpha }$
becomes $\exp \left[ -\frac{i\alpha }{2}\left( a^{2}+a^{\dag 2}\right) %
\right] $, which is still sort of one-mode squeezing operator. The kernel of
the new transformation is
\begin{equation}
\left\langle p\right\vert \tilde{K}_{\alpha }\mathfrak{M}^{\dag }\left\vert
x\right\rangle =\frac{\exp \left[ \frac{i}{2}\left( \frac{x^{2}+p^{2}}{\tanh
\alpha }-\frac{2xp}{\sinh \alpha }\right) \right] }{\sqrt{2\pi i\sinh \alpha
}},  \label{54}
\end{equation}%
note the similarity and difference between $\left( \tanh \alpha ,\sinh
\alpha \right) $ here in Eq. (\ref{54}) and $\left( \tan \alpha ,\sin \alpha
\right) $ in Eq. (\ref{1}).

Noticing the fact that Hermitian operator $a^{2}e^{i\theta }+e^{-i\theta
}a^{\dag 2}$ has no eigenstate, therefore the corresponding transformation $%
F_{\hat{K}_{\alpha }}$\ have no eigenfunction either. This conclusion
matches the fact that $\tilde{K}_{\alpha }$ is always sort of single-mode
squeezing operator, there should be no nontrivial eigenstate of $\tilde{K}%
_{\alpha }$.

In summary, by virtue of quantum mechanical representation transformation
and the method of integration within normal ordered product (IWOP) of
operators, we have found the key point of composing FrFT, and reveal the
structure for constructing new GFrT as the additive kernel-based linear
transformations, in so doing a full family of GFrT is discovered. Every
compositable kernel-based linear transformations can be \textquotedblleft
fractionalized" in our scheme. The eigen-functions of arbitrary GFrT are
derived explicitly also.

\end{document}